\documentclass[twocolumn,showpacs,preprintnumbers,draftclsnofoot]{revtex4}
\usepackage{amsmath}
\usepackage{amssymb}
\usepackage{graphicx}
\usepackage{dcolumn}
\usepackage{bm}
\usepackage{amssymb}
\usepackage{graphicx}
\usepackage{dcolumn}
\usepackage{bm}
\begin{document}

\title{ Engineering Majorana corner modes from two-dimensional hexagonal crystals   }
\author{Ma Luo\footnote{Corresponding author:luoma@gpnu.edu.cn} }%luoma@gpnu.edu.cn,swym231@163.com
\affiliation{School of Optoelectronic Engineering, Guangdong Polytechnic Normal University, Guangzhou 510665, China}

\begin{abstract}
Second order topological insulator can be engineered from two-dimensional materials with strong spin-orbit coupling and in-plane Zeeman field. In proximity to superconductor, topological superconducting phase could be induced in the two-dimensional materials, which host Majorana corner modes at the intersection between two zigzag edges. Two types of tight binding models in hexagonal lattice, which include $p_{z}$ or $p_{x,y}$ orbit(s) in each lattice site, are applied to engineer two-dimensional materials in topological superconducting phase. In both models, the condition that induces the second order topological superconductor requires nonuniform value of either in-plane Zeeman fields or superconductor pairing parameters in two sublattices. The finite size effect of the second model is weaker than that of the first model.

\end{abstract}

\maketitle

\section{Introduction}

Majorana zero mode (MZM) in condensed materials is analogy of Majorana Fermion in particle physics, which is anti-particle of itself \cite{Elliott2015,Aguado2017}. The MZMs can be potentially used to construct elementary quantum gates that are immuned to decoherence due to their physical nature of non-Abelian statistics \cite{Nayak08,Moore91,Wen91,Ivanov01,Nayak96,Sarma05}. Thus, topological quantum computing based on MZMs become feasible and attract tremendous attentions. Varying schemes have been proposed to realise the MZMs in one-dimensional or two-dimensional nano-structure, such as quantum wire \cite{Fidkowski11} or two-dimensional materials \cite{fu08,qi10,Akhmerov09,Law09,Chung11} in proximity to superconductor, and cold-atom systems \cite{Jiang11,Diehl11,Sato09,CZhang08,Tewari07,XJLiu14}. In one dimensional systems, the MZMs are localized at the end of the wire. In two dimensional system, the MZMs are localized at the core of vortex, which is induced by external magnetic field \cite{Sau10,Zhongbo20}. Another scheme to implement MZMs is to constructure heterostructures consisted of higher-order topological insulator and superconductor, which induces phase transition to the second order topological superconductors (SOTSCs) \cite{Langbehn17,QiyueWang18,Zhongbo18,TaoLiu18,Geier18,ChenHsuan18,Yanick19,Yuxuan18,Shapourian18,Zhigang19,SongBo20,Ghorashi19,RuiXing19,Plekhanov19,Skurativska20,Junyeong2020,Franca19,RuiXing19a,Bultinck19,YiTing20,Zhongbo19ab,Raditya20,Katharina20a,XiaoHong19,YaJieWu20,Xiaoyu18}.

One of the most intuitive scheme to implement MZMs is to construct heterostructure of two-dimensional second-order topological insulator (SOTI) in proximity to superconductor substrate \cite{Zhongbo19,YaJieWu20}. The two-dimensional SOTIs host corner modes at the corner that preserve the symmetric of the bulk crystals \cite{Benalcazar17,Benalcazar17a,TLi20,Schindler19,Benalcazar19,GMiert18}. In the absence of the proximity to superconductor, the corner modes of both electron and hole coexist. As the superconductor pairing parameters increase, the coupling between the electron and hole induces phase transition to the SOTSCs phase \cite{Zhongbo19,Zhongbo18}. In each pair of corner, there are only one pair of Majorana corner modes.

Although the minimal model of SOTI in two-dimensional square lattice is theoretically robust, the implementation in realistic materials is difficult, because the formation of two-dimensional systems is thermodynamically forbidden by the Mermin-Wagner theorem \cite{Mermin1966}. However, since the discovery of single sheet of graphene that violates the Mermin-Wagner theorem \cite{Xuekun1999,Xuekun1999a,QingTang2013}, multiple two-dimensional materials in hexagonal lattice have been discovered \cite{Sheneve2013,ChengChang2021}. Because of the advantage in mechanical and electronic properties of the two-dimensional materials, tremendous researches have been devoted to study their physics and application \cite{Castro09,Schaibley16}. Topological phases, such as $\mathbb{Z}_{2}$ topological insulator (TI) phase \cite{Kane05,Kane05a,Zhenhua11} and quantum anomalous Hall (QAH) phase \cite{Zhenhua10,Zhenhua14}, of two-dimensional hexagonal crystals (2D HCs) host robust edge states, which can be applied for high efficiency electronic devices \cite{Yong19,maluo19}. By applying in-plane Zeeman field in the $\mathbb{Z}_{2}$ TI phase of the 2D HCs, the helical edge states are gapped, which induce topological phase transition to the SOTI phase with robust corner modes at the the intersection between two zigzag edges \cite{YafeiRen2020}.

In this paper, we study the heterostructure consisted of 2D HCs that are intercalated between conventional superconductor substrate and ferromagnetic substrate. The superconductor substrate and the ferromagnetic substrate induce superconductor pairing and in-plane Zeeman field in the 2D HCs, respectively. We consider the general case that the superconductor pairing parameters or the Zeeman field at the two sublattices could be different. Two types of 2D HCs are considered. The first type is graphene-like materials. The tight binding model with one $p_{z}$ orbit in each lattice site is applied \cite{Castro09}. The intrinsic spin-orbit coupling (SOC) is weak \cite{Kane2005,Hongki2006,Yugui2007}, but could be enhanced by adatom doping \cite{Fufang2007,JunHu2012} or proximity effect\cite{Emmanuel2009,Jayakumar2014,Abdulrhman2018}. Our proposal of the MZMs is different from a previous proposal in Ref. \cite{XiaoHong19}, whose MZMs are localized at the intersection between zigzag edge and armchair edge of graphene. The second type includes materials such as arsenene \cite{SZhang18m,RGui19}, antimonene \cite{TZhou15,TZhou16,JJi16,GXu20}, bismuthene \cite{Reis17,TZhou18,ZSong14}, and binary element group-V monolayers \cite{ZLiu19m,SSLi17}, which is described by the tight binding model with $p_{x}$ and $p_{y}$ orbits in each lattice site \cite{TongZhou21}. For the materials in this type, the intrinsic SOC is strong, so that the TI phase is more feasible in experiment.

The paper is organized as follows: In Sec. II, the theoretical
model of the heterostructure are described. In Sec. III, the engineering of Majorana corner mode based on the first and second model is discussed. In Sec. IV, the conclusion is given.

\section{Theoretical model}

\begin{figure}[tbp]
\scalebox{0.58}{\includegraphics{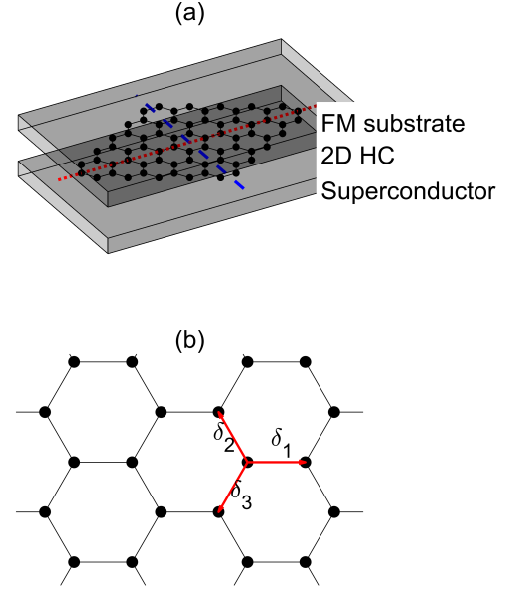}}% Here is how to import EPS art
\caption{ (a) Schematic of the two-dimensional hexagonal crystal (2D HC), which is sandwiched between the superconductor substrate and the ferromagnetic substrate. (b) Zoom in of the lattice structure of the 2D HC. The three directions of the nearest neighboring hopping are marked by $\delta_{1,2,3}$. }
\label{figure_system}
\end{figure}

The structure of the system is plotted in Fig. \ref{figure_system}(a). The 2D HCs is sandwiched between two substrates, which are s-wave superconductor and ferromagnetic insulator. The 2D HC is cut into rhombus flake with four zigzag edges. The two reflection symmetric axis are plotted as blue (dashed) and red (dotted) lines. Along each zigzag edge, there are $N_{zig}$ zigzag terminations. The top view of the 2D HC is plotted in Fig. \ref{figure_system}(b), which zoom in to show the detail of the lattice structure. The nearest neighboring pair of lattice is connected by three types of vectors, $\delta_{1,2,3}$, which are indicated by the red arrows.

In the absence of the superconductor, the carrier of the 2D HC is described by the tight binding Hamiltonian of electron. For the first model, the tight binding Hamiltonian is given as
\begin{eqnarray}
H=-t\sum_{\langle i,j\rangle,\sigma}c_{i,\sigma}^{\dag}c_{j,\sigma}+\frac{i\lambda_{I}}{3\sqrt{3}}\sum_{\langle\langle i,j\rangle\rangle}\sigma\nu_{ij}c_{i,\sigma}^{\dag}c_{j,\sigma} \nonumber \\
\sum_{i,\sigma,\sigma^{\prime}}(\lambda_{FM}+\eta_{i}\lambda_{AF})[\mathbf{s}\cdot\hat{M}]_{\sigma,\sigma^{\prime}}c_{i,\sigma}^{\dag}c_{i,\sigma^{\prime}}
\end{eqnarray}
, where $c_{i,\sigma}^{\dag}$ and $c_{i,\sigma}$ are creation and annihilation operator of electron of spin $\sigma=\pm1$ on the i-th lattice cite, $t$ is the strength of the nearest neighboring hopping, $\lambda_{I}$ is the strength of the intrinsic SOC, $\lambda_{FM}$ and $\lambda_{AF}$ are strength of the ferromagnetic and antiferromagnetic Zeeman field that are induced by the magnetic substrate, $\mathbf{s}=\sigma_{x}\hat{x}+\sigma_{y}\hat{y}+\sigma_{z}\hat{z}$ is the vector of Pauli matrix of spin, $\hat{M}$ is the unit vector that feature the direction of the Zeeman field. The first summation cover the nearest neighboring sites, and the second summation cover the next nearest neighboring sites. For the hopping between the next nearest neighboring sites i and j with left(right) turn, $\nu_{ij}=\pm1$. For the A(B) sublattice, $\eta_{i}=\pm1$. In this paper, we assume the parameters $t=2.8$ eV, and $\hat{M}=\hat{x}$.

For the second model, there are two orbits in each lattice site, which are $p_{x}$ and $p_{y}$ orbits. The tight binding Hamiltonian is given as
\begin{eqnarray}
H&=&-t\sum_{\mathbf{r}_{i},\mathbf{r}_{i}+\mathbf{\delta}_{j},\sigma}c_{\mathbf{r}_{i},\sigma}^{\dag}T_{\mathbf{\delta}_{j}}c_{\mathbf{r}_{i}+\mathbf{\delta}_{j},\sigma}
+\lambda_{I}\sum_{\mathbf{r}_{i},\sigma}\sigma c_{\mathbf{r}_{i},\sigma}^{\dag}\kappa_{z}c_{\mathbf{r}_{i},\sigma} \nonumber \\
&&+\sum_{\mathbf{r}_{i},\sigma,\sigma^{\prime}}(\lambda_{FM}+\eta_{\mathbf{r}_{i}}\lambda_{AF})[\mathbf{s}\cdot\hat{M}]_{\sigma,\sigma^{\prime}}c_{\mathbf{r}_{i},\sigma}^{\dag}c_{\mathbf{r}_{i},\sigma^{\prime}} \nonumber \\
&&+\lambda_{S}\sum_{\mathbf{r}_{i},\sigma}\eta_{\mathbf{r}_{i}}c_{\mathbf{r}_{i},\sigma}^{\dag}c_{\mathbf{r}_{i},\sigma}
\end{eqnarray}
, where $c_{\mathbf{r}_{i},\sigma}=[c_{\mathbf{r}_{i},\sigma}^{p_{x}},c_{\mathbf{r}_{i},\sigma}^{p_{y}}]^{T}$ is the column vector of annihilation operators of electron at the i-th site of spin $\sigma$ with each component being the operator of $p_{x}$ and $p_{y}$ orbits, $\mathbf{\delta}_{j}$ with $j=[1,2,3]$ are the three vectors connecting pairs of the nearest neighboring cites. The nearest neighboring hopping includes intra-orbit and inter-orbit hopping, which is described by the matrix
\begin{equation}
T_{\mathbf{\delta}_{j}}=\begin{bmatrix} t_{1} & z^{-(j+N_{\delta})}t_{2}\\
z^{j+N_{\delta}}t_{2} & t_{1}\end{bmatrix}
\end{equation}
, where $z=exp(2i\pi/3)$ and $N_{\delta}$ being integer. Varying value of $N_{\delta}$ is equivalent to rotating the lattice structure by $\pm120^{o}$, which is not effective to the bulk band structure, but have significant effect on the corner modes. $\kappa_{z}$ is the z-component Pauli matrix acting on the $p_{x,y}$ orbital space. For the A(B) sublattice, $\eta_{\mathbf{r}_{i}}=\pm1$. In addition to the intrinsic SOC and Zeeman field, the staggered sublattice potential with strength being $\lambda_{S}$ is included.

In the presence of superconductor substrate, the superconductor pairing is induced in the 2D HC flakes. By applying the Bogolyubov-de Gennes (BdG) theory, the effective Hamiltonian is given as
\begin{equation}
H_{BdG}=\begin{bmatrix}
H-\mu & H_{\Delta} \\
H_{\Delta}^{\dag} & -H^{T}+\mu
\end{bmatrix}\label{eq_BdGhamil}
\end{equation}
, where $\mu$ is the chemical potential, $H_{\Delta}=-i\sigma_{y}\Delta_{A}(1+\eta_{i})/2-i\sigma_{y}\Delta_{B}(1-\eta_{i})/2$ is the superconducting pairing with $\Delta_{A(B)}$ being the superconducting pairing parameter at the A(B) sublattice. In this paper, we consider only the case with $\mu=0$.

\section{Numerical result}

The quantum state of the 2D HC flake is obtained by diagonalization of the BdG Hamiltonian. In the absence of superconductor pairing and Zeeman field, the 2D HCs for both type of models with intrinsic SOC are $\mathbb{Z}_{2}$ topological insulator with helical edge state. In 2D HC flakes, the superposition between the clockwise and anticlockwise moving helical edge states form eigenstates of standing wave along the edge. In the presence of in-plane ferromagnetic Zeeman field, the time-reversal symmetry is broken, so that the helical edge state is gapped. Because the corner of two zigzag edges with intersecting angle being $120^{o}$ preserve the mirror-reflection symmetry of the bulk [reflection about the blue (dashed) line in Fig. \ref{figure_system}(a)], the system is in the topological crystalline phases, i.e. the SOTI phase. The zero energy corner states appear at the corner that preserve the symmetry \cite{YafeiRen2020}. At the $60^{o}$ corners, the mirror reflection does not exchange A/B sublattice, so that in the presence of $\lambda_{FM}$ the mirror reflection symmetry is not preserved. Thus, the $60^{o}$ corners do not host corner state. In the additional presence of superconductor pairing, the appearance of Majorana corner modes (MCMs) in the two models are discussed in the following two subsection.

\subsection{First model}

\begin{figure}[tbp]
\scalebox{0.58}{\includegraphics{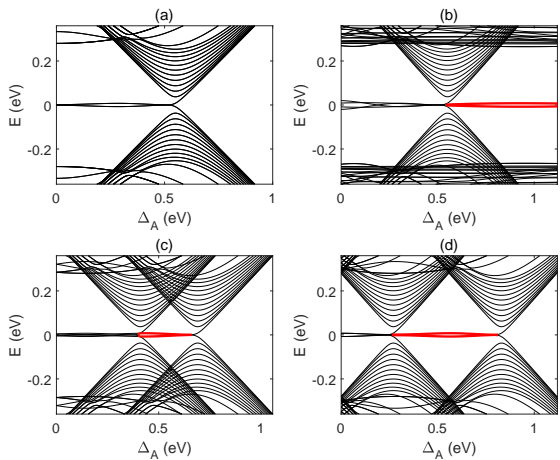}}% Here is how to import EPS art
\caption{ Energy spectrum of the 2D HC flakes described by the first model versus the superconductor pairing parameters. In (a,c,d), $\Delta_{A}=\Delta_{B}$; in (b), $\Delta_{B}=0.25$ eV with varying $\Delta_{A}$. In (a,b), $\lambda_{AF}=0$; in (c), $\lambda_{AF}=0.05t$; in (d), $\lambda_{AF}=0.1t$. The other parameters are $\lambda_{I}=0.1t$ and $\lambda_{FM}=0.2t$. The thick red lines mark the spectrum of the MCM. }
\label{figure_band1}
\end{figure}

For the first model, we consider the 2D HC flake with $N_{zig}=17$, $\lambda_{I}=0.1t$ and $\lambda_{FM}=0.2t$. If both of the superconductor pairing and the Zeeman field are uniform, i.e. $\Delta_{A}=\Delta_{B}$ and $\lambda_{AF}=0$, the energy spectrum of the 2D HC flake versus the superconductor pairing is plotted in Fig. \ref{figure_band1}(a). As the strength of the superconductor pairing increase, the gaps of bulk states and helical edge states are both decreased. As $\Delta_{A(B)}$ reaches the value of $\lambda_{FM}$, the band gap of the helical edge states become zero, so that the energy spectrum of the 2D HC flake become semi-continue at zero energy. As $\Delta_{A(B)}$ further increase, the band gap is reopened. In this case, the 2D HC flake is topologically trivial, so that there is not any corner mode. In Fig. \ref{figure_band1}(b), a different case with nonuniform superconductor pairing, i.e. $\Delta_{A}\ne\Delta_{B}$, is considered. As the condition of either $\Delta_{A}=\lambda_{FM}$ or $\Delta_{B}=\lambda_{FM}$ being satisfied, the band gap of the helical edge states closes and reopened. In the phase regime with $\Delta_{A}<\lambda_{FM}$ and $\Delta_{B}>\lambda_{FM}$ ($\Delta_{A}>\lambda_{FM}$ and $\Delta_{B}<\lambda_{FM}$), the 2D HC is in the SOTSC phase. Two MCMs appear at the top and bottom corners, so that there is averagely one MCM at each corner.

\begin{figure}[tbp]
\scalebox{0.58}{\includegraphics{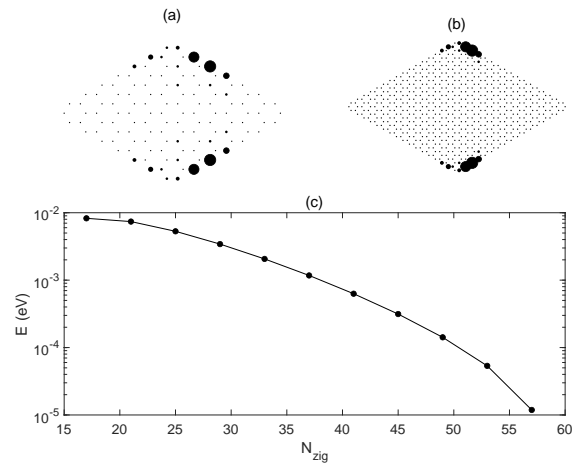}}% Here is how to import EPS art
\caption{ (a) Spatial distribution of the mode amplitude for the 2D HC flake described by the first model with $N_{zig}=7$, $\lambda_{I}=0.1t$, $\lambda_{FM}=0.2t$, $\lambda_{AF}=0.1t$, and $\Delta_{A}=\Delta_{B}=0.6$ eV. (b) The same as (a) except $N_{zig}=17$. (c) Energy level of the MCM versus $N_{zig}$. }
\label{figure_wave1}
\end{figure}

Since the nonuniform superconductor pairing is difficult for experimental implementation, another scheme with uniform superconductor pairing and nonuniform Zeeman field is considered. In this case, both $\lambda_{FM}$ and $\lambda_{AF}$ are nonzero. With $|\lambda_{AF}|<|\lambda_{FM}|$, the Zeeman field is ferrimagnetic instead of ferromagnetic, which could be engineered by proximity effect \cite{Petra20,Sebastian21}. The energy spectrum of the 2D HC flake with $\lambda_{AF}=0.05t$ is plotted in Fig. \ref{figure_band1}(c). The gap of the helical edge states closes with the condition of either $\Delta_{A(B)}=\lambda_{FM}-\lambda_{AF}$ or $\Delta_{A(B)}=\lambda_{FM}+\lambda_{AF}$. As $\Delta_{A(B)}<\lambda_{FM}-\lambda_{AF}$, the 2D HC flake remains in the SOTI phase with four corner modes. As $\lambda_{FM}-\lambda_{AF}<\Delta_{A(B)}<\lambda_{FM}+\lambda_{AF}$, the 2D HC flake is in the SOTSC phase with two MCMs. For the 2D HC flake with $\lambda_{AF}$ being larger, the phase regime of SOTSC phase become wider, as shown in Fig. \ref{figure_band1}(d).

The presence of the topological corner states at the intersection between two zigzag edges is due to the inversion of the winding numbers of the one-dimensional edge bands at the two edges. With $\Delta_{A(B)}=0$, $\lambda_{AF}=0$, and $\lambda_{FM}\ne0$, the band inversion of the edge states of both electrons and holes coexist, so that the corner states of electron and hole coexist. In the presence of small $\Delta_{A(B)}$ in the SOTI phase, the edge states of electrons and holes in each zigzag edge weakly couple into two mixed edge bands, which does not change the winding numbers. Thus, the Majorana Kramers pairs of zero modes appear at each $120^{o}$ corner. As $\Delta_{A}$ and $\Delta_{B}$ both increase and exceed $\lambda_{FM}$ under the condition $\Delta_{A}=\Delta_{B}$ , the band gap of the edge states close and reopen, which turn the winding number of the two mixed edge bands into zero. Thus, there is not corner state. If $\Delta_{A}$ and $\Delta_{B}$ vary under the condition $\Delta_{A}\ne\Delta_{B}$, the band gap of only one of the two mixed edge bands closes and reopens at the condition $\Delta_{A(B)}=\lambda_{FM}$. Similarly, if $\Delta_{A}$ and $\Delta_{B}$ vary under the condition $\Delta_{A}=\Delta_{B}$ but $\lambda_{AF}\ne0$, only one of the two mixed edge bands closes and reopens at the condition $\Delta_{A(B)}=\lambda_{FM}\pm\lambda_{AF}$. In these two cases, the winding number of one of the two mixed edge bands becomes zero, and that of the other edge band remains being $\pm1$. Thus, there is only one corner mode at the $120^{o}$ corner, which is the MZM.

For a particular 2D HC in the SOTSC phase, the spatial distribution of the mode amplitude of the MCM are plotted in Fig. \ref{figure_wave1}(a) and (b). The mode pattern mainly distributes at two zigzag terminations to the right side of the top and bottom corners. The zigzag terminations at the two zigzag edges to the left(right) of the vertical axis in the middle belong to the A(B) sublattice. Thus, the localization of the mode pattern is determined by the sign of $\lambda_{AF}$. As $\lambda_{AF}=-0.1t$ present, the mode pattern mainly distributes at two zigzag termination to the left side of the top and bottom corner, which is mirror reflection of those in Fig. \ref{figure_wave1}(a) and (b) about the middle axis. In Fig. \ref{figure_wave1}(a), the size of the flake is small, so that the MCMs at the top and bottom corner have sizable overlap. Thus, the energy level of the MCMs are not zero but a finite small number. As the size of the 2D HC flake become larger, the mode pattern at the top and bottom corners have smaller overlap, as shown in Fig. \ref{figure_wave1}(b), so that the energy level is nearer to zero. The energy level of the MCMs versus the size of the 2D HC flake is plotted in Fig. \ref{figure_wave1}(c), which exhibits exponential decay. However, the decay rate is slow, due to the weak localization of the MCMs at the corners.

\subsection{Second model}

\begin{figure}[tbp]
\scalebox{0.58}{\includegraphics{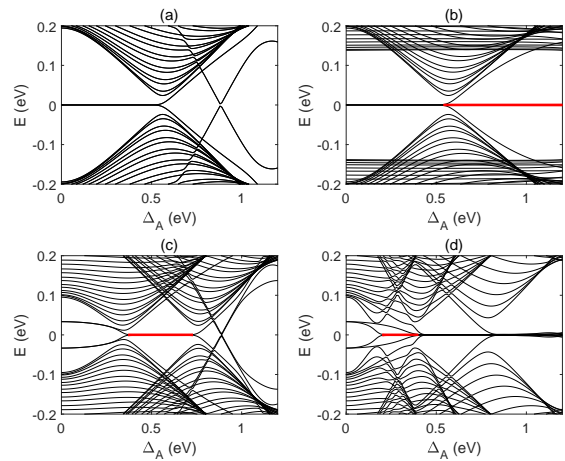}}% Here is how to import EPS art
\caption{ Energy spectrum of the 2D HC flakes described by the second model versus the superconductor pairing parameters. In (a,c,d), $\Delta_{A}=\Delta_{B}$; in (b), $\Delta_{B}=0.25$ eV with varying $\Delta_{A}$. In (a,b), $\lambda_{AF}=0$; in (c,d), $\lambda_{AF}=0.1t$. In (a,b,c), $N_{\delta}=2$; in (d), $N_{\delta}=3$. The other parameters are $\lambda_{I}=0.3t$, $\lambda_{S}=0.1t$, and $\lambda_{FM}=0.3t$. The thick red lines mark the spectrum of the MCM. }
\label{figure_band2}
\end{figure}

\begin{figure}[tbp]
\scalebox{0.58}{\includegraphics{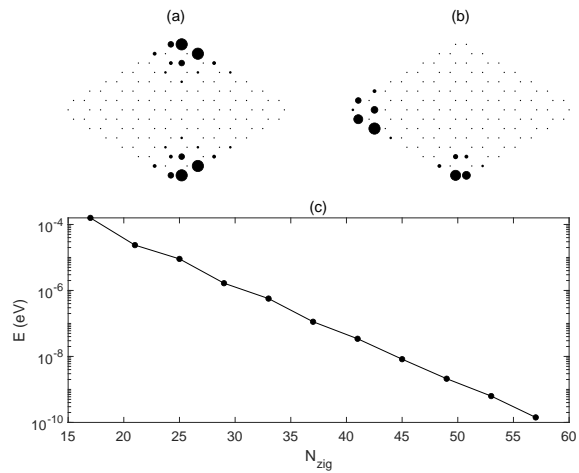}}% Here is how to import EPS art
\caption{ (a) Spatial distribution of the mode amplitude for the 2D HC flake described by the second model with $N_{zig}=7$, $N_{\delta}=2$, $\lambda_{I}=0.3t$, $\lambda_{S}=0.1t$, $\lambda_{FM}=0.3t$, $\lambda_{AF}=0.1t$, and $\Delta_{A}=\Delta_{B}=0.6$ eV. (b) The same as (a) except $N_{\delta}=3$ and $\Delta_{A}=\Delta_{B}=0.3$ eV. (c) Energy level of the MCM of the flake in (a) versus $N_{zig}$. }
\label{figure_wave2}
\end{figure}

For the second model, the 2D HC flake with the same size of those in Fig. \ref{figure_band1} are considered. The other parameters are $t_{1}=t_{2}=t=1$ eV, $\lambda_{I}=0.3t$, $\lambda_{FM}=0.3t$, and $\lambda_{S}=0.1t$, which are typical value for the related materials. Although the lattice structure have $C_{3}$ rotational symmetric, the phase factor of the inter-orbit hopping in $T_{\delta_{j}}$ breaks the rotational symmetric. As a result, the 2D HC flakes with different value of $N_{\delta}$ have different phase diagram. As $N_{\delta}=2$, the mirror reflection symmetric with vertical axis in the middle is preserved, so that we studied this case firstly. For the other two types of 2D HC flakes with $N_{\delta}=1$ or $N_{\delta}=3$, the mirror reflection transfer one type of model to the other type, so that they have the same energy spectrum and topological properties.

In the 2D HC with $N_{\delta}=2$, uniform superconductor pairing and uniform Zeeman field, the energy spectrum is plotted in Fig. \ref{figure_band2}(a), which does not have MCM. As $\Delta_{A(B)}<\lambda_{FM}$, the 2D HC flake is in the SOTI phase with four corner states at the top and bottom corners. As $\Delta_{A(B)}>\lambda_{FM}$, two second order degenerated corner modes with changing energy levels for varying $\Delta_{A(B)}$ appear, which are localized at the left and right corners. As the superconductor pairing become nonuniform, the SOTSC phase appear in the phase regime with $\Delta_{A}<\lambda_{FM}$ and $\Delta_{B}>\lambda_{FM}$ ($\Delta_{A}>\lambda_{FM}$ and $\Delta_{B}<\lambda_{FM}$). A typical example is plotted in Fig. \ref{figure_band2}(b) with $\Delta_{B}=0.25$ eV and varying $\Delta_{A}$. In another scheme that the superconductor pairing is uniform, but $\lambda_{AF}$ is nonzero, the energy spectrum is plotted in Fig. \ref{figure_band2}(c). In the absence of superconductor (i.e. $\Delta_{A(B)}=0$), the presence of the antiferromagnetic Zeeman field break the SOHI phase, so that the energy levels of the corner modes is nonzero. As $\Delta_{A(B)}$ increase, the energy levels of the corner modes as well as the band gaps of the helical edge states decrease. As $\Delta_{A(B)}$ exceed the condition $\Delta_{A(B)}=\lambda_{FM}-\lambda_{AF}$, the band gap closes and reopens, which trigger phase transition to the SOTSC phase with two MCMs. The phase regime of the SOTSC phase is given by the condition $\lambda_{FM}-\lambda_{AF}<\Delta_{A(B)}<\lambda_{FM}+\lambda_{AF}$. For a particular 2D HC flake with $\Delta_{A(B)}=0.6t$, the spatial distribution of the mode pattern of the MCM is plotted in Fig. \ref{figure_wave2}(a). The mode pattern mainly localized at the top and bottom corners and one zigzag termination to the right of the corners. Comparing to the mode pattern of the MCM of the first model in Fig. \ref{figure_wave1}(a,b), the mode pattern of the second model has higher degree of localization, so that the finite size effect is weaker. The inference is confirmed by plotting the energy level of the MCM versus the size of the 2D HC flake, as shown in Fig. \ref{figure_wave2}(c). The energy level exponentially decay to $10^{-10}$ eV as the size reaches $N_{zig}=57$.

In another case with the same parameters as those in Fig. \ref{figure_band2}(c) except $N_{\delta}=3$, the energy spectrum versus $\Delta_{A(B)}$ is plotted in Fig. \ref{figure_band2}(d).  The phase regime of SOTSC becomes a narrow range of $\Delta_{A(B)}$ near to $\lambda_{FM}-\lambda_{AF}$, as shown by the red lines in Fig. \ref{figure_band2}(d). When $\Delta_{A(B)}=0.3$ eV, there are two MCMs, which are localized at the bottom and left corners, as shown by the spatial distribution of the mode pattern in Fig. \ref{figure_wave2}(b). If the sign of $\lambda_{AF}$ is flipped to be negative, the MCMs are localized at the top and right corners. As $\Delta_{A(B)}$ become larger (away from the red curve in Fig. \ref{figure_band2}(d)) but smaller than $\lambda_{FM}+\lambda_{AF}$, there are six corner states in total, which have nearly zero energy. The four corner states at the top and right corners mix with the two MCMs. In this case, the 2D HC flake is in the intermedius phase between SOTSC and SOHI. As $\Delta_{A(B)}>\lambda_{FM}+\lambda_{AF}$, there are eight corner states in total, and averagely two corner states at each corner, so that the 2D HC flake is in the SOHI phase.

\section{Conclusion}

In conclusion, combination of superconductor pairing and in-plane ferrimagnetic Zeeman field in 2D HCs could induce SOTSC phase, which host MCM at the corner between two zigzag edges. Although the two models of 2D HCs can both become SOTSC, the second model is more superior due to two reasons: the finite size effect is smaller; the large intrinsic SOC is originated from the atoms of pristine 2D HCs without doping or proximity effect. As a result, the MCMs in the 2D HC flakes of the second model is feasible for experimental implementation.

\begin{acknowledgments}
This project is supported by the Natural Science Foundation of Guangdong Province of China (Grant No.
2022A1515011578), the Project of Educational Commission of Guangdong Province of China (Grant No. 2021KTSCX064), the startup grant at Guangdong Polytechnic Normal University (Grant No. 2021SDKYA117), and the National Natural Science Foundation of China (Grant No.
11704419).
\end{acknowledgments}

\clearpage

\end{document}